\documentclass[floatfix, preprintnumbers,aps, amsmath, amssymb, showpacs, 10pt, ]{revtex4}
\usepackage{graphicx}
\usepackage{dcolumn}
\usepackage{epsfig}
\usepackage{bm}
\begin{document}

\date{\today}

\title{Structured FRW universe leads to acceleration: a non-perturbative approach}

\author { Reza Mansouri} \email[On sabbatical leave from Department 
of Physics, Sharif University of Technology, Tehran, and Institute
 for Studies in Physics and Mathematics(IPM), Tehran.]{mansouri@hep.physics.mcgill.ca; mansouri@ipm.ir}
 
\affiliation{Dept. of Physics, McGill University, 3600 University Street, Montreal QC,\\
 H3A 2T8, Canada}

\begin{abstract}
We propose a model universe in the matter dominated phase described by a FRW background
with local inhomogeneities, like our local patch, grown out of the primordial 
fluctuations. Our sub-horizon local patch consisting of different structures is approximated 
as an inhomogeneous cosmic fluid described by a LTB metric embedded in a 
background FRW universe, in which the observer could be located anywhere. Within the exact general relativistic formulation, the 
junction conditions for the only possible matching without a thin shell at the 
boundary, neglected so far in the literature, constrains the model in such a way 
that the luminosity distance-red shift relation mimics a FRW universe with dark 
energy. Therefore, the dimming of SNIa is accounted for in such a 
{\it structured} FRW universe. We have also calculated the exact general relativistic backreaction term and shown how it influences the global Hubble parameter and the effective density of the cosmic fluid
By using an exact formulation of the general relativistic dynamics of 
structures in a homogeneous universe, the claim is therefore stressed that the backreaction of cosmological perturbations leads to an apparent dimming of the cosmological distances.  
\end{abstract}


\maketitle

\newcommand{\eq}[2]{\begin{equation}\label{#1}{#2}\end{equation}}
 
\section{Introduction}

 The Copernican turn in cosmology, which happened after the identification 
of two different population of Cepheids in the mid fifties of the last 
century, was a milestone for the acceptance of homogeneity of the
universe and the FRW metric as {\it the metric} of our universe. Although 
the inhomogeneity of the local structure of the universe had been 
observed, but all the observational data had led to the acceptance 
of the homogeneity at scales larger than some hundred mega parsecs\cite{Peebles, Lahav},
 and its generalization to all of the universe. The introduction of a homogeneous cosmic
fluid describing the matter content of the universe was the natural theoretical formulation 
of this observational finding. So far there has been excellent agreement between theory and
observation within the limit of observational precision. That is why FRW universe had been 
accepted as {\it the model universe} on which the interpretation of all the observational 
data is based, albeit many observations in the last decade show us explicitly the inhomogeneity
of the cosmic structures on the cosmological scales of our surrounding.  \\ 
 The precision cosmology is, however, already so far developed that we can not ignore any 
more the local inhomogeneity of the universe, and the theoretical concept of the homogeneous 
cosmic fluid, i.e. the basic concept of the FRW models, has to be modified. On the other 
hand, the well established successes of standard FRW cosmology can not be abandoned so 
easily. \\
 The deviation from the standard homogeneous cosmic fluid in the matter dominated phase 
of the universe in our proximity should be reflected in the data from cosmological 
objects. Now, recent observational data on SNIa imply a larger distance to supernovae
than predicted by the conventional FRW universe \cite{Riess 98, Perlmutter 99}. 
Different factors, such as evolution effects, dust absorption, gravitational
lensing, or dynamics of the universe, may have led to this dimming of 
SNIa's. Detailed studies show that all these factors have negligible 
effects, except the dynamics, which had led to the "term acceleration 
of the universe" as the sole model-independent interpretation of the 
data \cite{Riess 04, Fillipenko 04}. Of course, the familiar interpretation
is within the context of the standard FRW cosmology, as all the theoretical 
ingredients, such as cosmological constant and all the conceivable equation 
of states derivable from a scalar field been developed before. Hence, the 
concept of dark energy, in addition to the baryonic and dark matter content 
of the universe, for the interpretation of the dimming of SNIa within the 
FRW universe is still the prevailing hypotheses. \\
Since then, papers related to dark energy are increasing in torrent, many different 
models have been developed, and many terms related to it like quintessense, 
k-essence, spin-essence, phantom, mirage, and so on have been coined in the last 
years, many of them violating basic physics intuition. Leaving aside modified 
gravity theories in 4 dimensional spacetimes, and many different models of 
brane cosmologies, there are two conventional approaches to explain the 
acceleration of the universe: back-reaction of cosmological perturbations and 
the inhomogeneous models of the universe. Although these two approaches use 
different methodologies and techniques of calculations, they are based on similar 
assumptions as far as the homogeneity of the universe is concerned, as it 
is best elaborated in\cite{Kolb 05}. In the back-reaction approach 
one assumes a sub- or super-horizon cosmological perturbation and tries to  
account for the acceleration of the universe using different perturbative 
approaches\cite{Geshniz 02,  bran 02, Martineau 05, KMNR 05, rasanen, notari, KMR 05}. 
It has been shown that both sub-\cite{Kolb 05} and super-horizon
\cite{Martineau 05} perturbations may lead, at least partially, to an 
explanation of the acceleration of the universe without invoking dark 
energy or modifying gravity. However, the perturbative methods used in 
this approach leave enough space for criticism\cite{Geshniz 05, hirata 05, Flanagan 05}.\\
Incorporating the large scale structures of the universe into a model 
universe has led different authors to look at the inhomogeneous models
and its consequences for the cosmological parameters. Even long before the 
new SNIa data, the possibility of distinguishing observationally the 
homogeneous from inhomogeneous past light-cone had been investigated by
Partovi and Mashhoon\cite{partovi}. Just after the release of the new data 
on SNIa, Celerier\cite{celerier} published an interesting paper questioning
the dark energy interpretation of the acceleration of the universe. She
showed, using a Lemaitre-Tolman-Bondi(LTB) inhomogeneous solution of 
the Einstein equations\cite{LTB} and the corresponding luminosity distance 
relation in it, that large scale inhomogeneities may mimic a cosmological 
constant, say dark energy, in a homogeneous universe. Tomita\cite{Tomita 00, 
Tomita 01}, in an attempt to explain the acceleration, uses an 
inhomogeneous model universe consisting of an underdense FRW region, a 
void of about 200 Mpc extend, immersed in the FRW bulk, and saw hints of 
acceleration due to the underdensity of the void, without answering the 
question of the effect of the thin shell supporting energy and momentum needed 
at the boundary of the bulk. Giovannini\cite{giovanni}, along the same line 
as the authors of\cite{hirata 05, Flanagan 05}, assuming an arbitrary inhomogeneous 
metric, shows that- within the limit of some approximations- the deceleration 
parameter of a matter dominated universe is always positive. The {\it no-go 
theorem} adopted in these references is repeated in other papers dealing 
with a LTB inhomogeneous universe. Govannini then concludes that the claimed 
acceleration must be the result of extrapolation of a specific solution 
in a regime where both the perturbative expansion breaks down and the 
constraints are violated. Wiltshire\cite{wiltshire 05}, generalizing  
Tomita's model, assumes an inhomogeneous underdense model within a bulk 
FRW universe, without going into the dynamics of matching of two 
different solution of the Einstein equations and finds 'promising' 
consequences. Moffat\cite{moffat}also assumes an inhomogeneous LTB void 
within a FRW universe where the matching is placed at about $z = 20$, not 
going into the detail of the dynamics of Einstein equation for such a 
matching. The result is again the possibility of explaining the acceleration 
without any need of a dark energy. Alnes et. al.\cite{alnes}, assuming again a 
model of an inhomogeneous void immersed smoothly in a FRW universe, go a step 
further considering different models according to the distribution of the 
density in the void and looking at their cosmological consequences such 
as luminosity distance relation, and the position of the first CMB peak. 
They too, ignoring the details of the dynamics of the Einstein equations for such a matching, come to the conclusion that in some of the models proposed it maybe possible to explain the acceleration without any use of a dark energy. Bolejko\cite{bolejko}, using again a void embedded in a FRW universe, and assuming 6 different models for the density distribution, comes to the result that there is no realistic model which could explain the observed dimming of supernovae without a cosmological constant, not noting the impossibility of such a junction. \\
Models of an inhomogeneous bubble embedded in a FRW have the advantage of
admitting, in principle, an exact approach, remedying the main shortcoming 
of the perturbative back-reaction approach as has been mentioned by many 
authors. But, as mentioned before, all the papers published so far neglect 
one important point: is it at all possible, within general relativity, to 
have an underdense spherical bubble embedded in a bulk FRW universe? A 
negative answer would catapult all these analyses into the range of 
approximative approaches with the result that all the criticism published 
so far on the back-reaction approaches will apply to these models too.\\
We have analyzed this question in detail some years ago\cite{khak}. It has 
been shown there explicitly that the matching of an underdense LTB bubble
to a FRW universe is not possible, except for the case of having a thin 
shell supporting energy and momentum on the boundary to the FRW background.
Apart from lacking astrophysical indications of such a spherical
thin mass condensation around us at about $z = 0.46$, i.e. the boundary 
of transition from the accelerating to decelerating epochs of the 
universe\cite{riess 04}, we do not know of any other theoretical indication
for it from the large scale studies of the universe. On the contrary, an 
overdense region surrounded by a void as the result of the evolution of the 
primordial perturbation is the most expected one. The case of the great wall being 
considered as a massive thin shell around a putative local void\cite{zehavi}, even 
if it turns out to be observationally viable, is just a perturbation within the cosmic
fluid of our local patch we are going to consider which extends up to about $z = 0.46$  \\ 
 Here we report on a realistic exact GR model of the universe, consisting of 
inhomogeneous patches embedded smoothly in a FRW background without any thin
shell to be required. We suggest minimal changes to the FRW universe
incorporating the inhomogeneity in our cosmic neighborhood. The theoretical 
description of the real inhomogeneity of the structures in our proximity is modelled
again by a cosmic fluid which, contrary to the FRW case, is inhomogeneous. We will
have to constrain the infinite number of degrees of freedom of such a cosmic fluid
model to the smallest possible number to match the observations, otherwise the complexity
of the model makes it useless for observational cosmology. Therefore, our main task is
to provide a more realistic model universe capable of matching the precision cosmology
of today and its future developments. Much of the work done so far within the FRW 
models have to be repeated now to incorporate the inhomogeneity of the local patches
and provide a new theoretical framework for interpretation of the observational data
The {\it structured }FRW (SFRW) model we are proposing is to be considered as a first step in that 
direction.  \\  
The local patches grown out of the primordial perturbations and their backreactions
to the homogeneous background are modelled exactly as a truncated flat LTB manifold 
embedded in a FRW universe from which a sphere of the same extent as the LTB patch 
is removed. It turns out that as a result of the junction conditions the mean density 
of any such inhomogeneous patch, with over- and under-dense regions, has to be equal to 
the density of the FRW bulk. Therefore, the Copernican principle is in no way 
violated and we are led to a structured universe where the local patches are 
distributed homogeneously in the bulk and having the same mass 
as a local FRW patch would have, accounting for all the structures we see grown out
of the primordial perturbation within a FRW universe. The analysis of the luminosity 
distance relation in our structured FRW model shows explicitly a dimming of objects 
within a patch relative to what it would be inferred from a standard FRW universe. 
The so-called 'bang time function'\cite{celerier, bolejko, hellaby}, which is an 
integration function in LTB bubble models, is interpreted very naturally 
as the time of nucleation of mass condensation in a patch and its behavior 
is fixed through the junction conditions at the transition epoch.\\
\paragraph{Symbols and Notations:} 
 
 The index $n$ indicates the nucleation time, and index $b$ is 
for the boundary to the FRW universe; Because of the homogeneity of FRW it could be 
also read as the bulk. The index $c$ at $\rho_c$ indicates the density with respect 
to the specially defined comoving coordinate. Time $t$ may denote the coordinate or
the time at the observer, easily distinguished in the context. The index $0$ refers
to quantities at the observer time and coordinate, i.e. at our vicinity. By peculiar 
velocity we always mean radial peculiar velocity, to be distinguished from the general 
usage in astrophysics. $\Sigma$ is the boundary of our local patch to the FRW background.  

\section{The structured FRW (SFRW) model}
The assumed homogeneity of the universe is at the scales greater than some 
hundred mega parsecs. In regions below that we have different structures showing
the inhomogeneity in the smaller scales. Up to know, we have always interpreted 
the astrophysical data, at any scale whatsoever, on the basis of the assumption
that the matter content of the universe is best modelled through the 
homogeneous cosmic fluid, which is achieved over some large smoothing 
scale\cite{Lahav}, and always tacitly assumed that the fine grained details are 
smoothed out and ignored.  The simplicity of FRW universe, reducing the 
infinite degrees of freedom of the real universe to just one scale 
factor, has been the compelling reason for all the data interpretations so far.\\ 
Now, let us go a step back in the smoothing process and make our model more 
realistic to see if any substantial differences in the interpretation of data may 
results. We remove a spherical patch, resembling our local neighborhood in the 
universe up to about $z = 0.5$\cite{riess 04}, out of the FRW universe model and replace it by a simple inhomogeneous spherical mass distribution. The simplest way of modeling our local patch is to use a LTB flat metric, without any cosmological 
parameter. Our local patch is embedded in a flat FRW universe.
This does not contradict the cosmological principle, nor is it a reaction 
to Copernican turn, as the universe is full of different patches like ours
distributed homogeneously in the background FRW due to the existence of 
primordial density fluctuations. Therefore, the result of interpretation of 
the observed large scale data maybe the same in any other patch within the 
FRW universe. As our local patch is the consequence of a primordial perturbation, or
mass condensation, within a FRW background, it must be matched to a FRW metric smoothly.
Otherwise, we have either to change our gravity theory, abandon the exactness of our 
calculation and accept the perturbative nature of it, or assume a thin mass 
condensation at the boundary of our local patch to the FRW and looking for a 
mass deficit or surplus in the patch in comparison to the mass density of the 
background universe. Non of these alternatives is desired or observed. Having 
this model in mind, we look for the exact dynamic of such a model. Note that 
we are defining in principle a FRW universe having structured patches 
within it distributed homogeneously and isotropically, although each patch 
is inhomogeneous. Our SFRW model could be considered as a generalization of the idea of the Swiss cheese model, in which the subhorizon inhomogeneous patches are distributed homogeneously and we are living somewhere in one the patches. The model is exact in the sense of being an exact solution of the Einstein equations. 

\subsection{Dynamics of a patch within the structured FRW universe} 
   
Our local inhomogeneous matter dominated patch is modelled as an inhomogeneous 
spherically symmetric manifold of comoving radius $r_b = L$ with an arbitrary density profile 
glued to a homogeneous pressure-free FRW background from which a sphere of matter 
of the same radius is removed. Our calculation is based on an exact general 
relativistic formulation of gluing manifolds. This may be considered 
as a generalization of the work done by Olson and Silk\cite{silk} 
within the Newtonian dynamics where there is no need to be cautious about 
the matching conditions.\\
 According to a theorem in general relativity, there is no solution of 
Einstein equations representing a time-dependent fluid sphere with finite 
radius having an equation of state in the form $\rho = \rho(p)$\cite{mansouri}. 
Inhomogeneous dust fluid defined by $p = 0$ representing the matter dominated 
phase of our local patch of the universe does not violate this theorem, contrary 
to the radiation dominated phase defined by the equation of state $\rho = 3p$. 
Therefore, we may continue with our model, and take a finite 
matter dominated patch of the universe represented by an inhomogeneous dust 
cosmic fluid obeying $p = 0$.\\
 Our spherical inhomogeneous patch containing dust matter is represented by a 
LTB metric embedded in a pressure-free FRW background universe with the 
uniform density $\rho_{b}$. We choose the LTB metric to be written in the 
synchronous comoving coordinates in the form\cite{khak}:
\begin{eqnarray}\label{metric}
ds^{2} = -dt^{2}+\frac{R'^{2}}{1+2E(r)}dr^{2}+R^{2}(r,t)(d\theta^{2}+\sin^{2}
\theta d\phi^{2}).
\end{eqnarray}
 The overdot and prime will thereafter denote partial differentiation with respect to $t$
 and $r$, respectively, and $E(r)$ is an arbitrary real function such that
 $E(r) > - \frac{1}{2}$. Then the corresponding Einstein equations turn out to be
\begin{eqnarray}
&&\dot{R}^{2}(r,t) = 2E(r)+\frac{2M(r)}{R} ,\\
&&\hspace*{0.6cm}4\pi\rho(r,t) = \frac{M'(r)}{R^2 R'}.
\end{eqnarray}
The density $\rho (r,t)$ is in general an arbitrary function of $r$ and $t$, and the 
integration time-independent function $M(r)$ is defined by
\begin{equation}
M(r) = 4\pi \int^{R(r,t)}_{0}\rho(r,t)R^{2}dR
     = \frac{4 \pi}{3} \overline{\rho}(r,t) R^3, 
\end{equation}
where $\overline {\rho}$, as a function of $r$ and $t$, is the average density 
up to the radius $R(r,t)$. \\
Furthermore, in order to avoid shell crossing of dust matter during their
radial motion, we must have $R'(r,t)>0$. Solutions to the above equations show
that an overdense spherical inhomogeneity with $E(r)<0$ within $R$ evolves just like
a closed universe, namely it reaches to a maximum radius at a certain time, then the
expansion ceases and undergoes a gravitational collapse so that a bound object forms
in such a way. In other words, $E(r)$ plays the role of the curvature scalar $k$ in 
the FRW universe.\\
For the sake of simplicity and comparison to the astrophysical parameters, we take the 
solution of the dynamical equation (2) which corresponds to $E(r) = 0$, the so-called
flat or parabolic case. The solution can be written in the form\cite{celerier, bolejko, hellaby}:
\begin{equation}
R(r,t) = (\frac{9M(r)}{2})^{\frac{1}{3}} (t -t_n (r))^{\frac{2}{3}},
\end{equation}
where $t_n (r)$ is an arbitrary function of $r$ appearing as an integration 'constant'.
This arbitrary function has puzzled different authors who give it the name of 'bang time
function' corresponding to the big bang singularity\cite{bolejko, celerier, hellaby}. It 
has, however, a simple astrophysical meaning within our structured FRW universe.
As $R(r,t)$ is playing the role of radius of our local patch, the time
$t = t_n$, leading to $R = 0$, means the time of onset of the mass condensation or 
nucleation within the homogeneous cosmic fluid. That is why we prefer to use 
the subscript $n$ for it indicating the time of nucleation. In the next section
we will see its crucial role in the luminosity distance relation and the impact of
the junction conditions on its running.\\ 

The metric(\ref{metric}) can also be written in a form similar to the Robertson-Walker metric. The definition
$$a(t, r) = \frac{R(t, r)}{r},\hspace{1cm} k(r) = - \frac{2E(r)}{r^2}$$
brings the metric into the form
\begin{equation}
ds^2 = -dt^2 + a^2\big[\big(1 + \frac{a' r}{a}\big)^2 \frac{dr^2}{1-k(r)r^2}+
       r^2d\Omega^2 \big].
\end{equation}
For a homogeneous universe, $a$ and $k$ don't depend on $r$ and we get the familiar Robertson-Walker metric. In our SFRW universe, the metric outside the inhomogeneous patch, is Robertson-Walker again.  \\
The corresponding field equations and the solution for the parabolic case $E(r) = 0$ can be written in the following familiar form:
\begin{equation}
\big (\frac{\dot a}{a} \big) = {1\over 3} \frac{\rho_c(r)}{a^3} - \frac{k}{a^2},
\end{equation}
where we have introduced $\rho_c(r) \equiv \frac{6M(r)}{r^3}$. These are very similar to the familiar Friedmann equations, except for the $r$-dependence of the different quantities. The solution (5) for the parabolic case can now be written in the form:
 \begin{equation}
a(r) = (\frac{3}{4}\rho_c(r))^{1\over 3}(t- t_n(r))^{2\over3}. 
\end{equation}
Now, let us denote by $\Sigma$ the (2+1)-dimensional timelike boundary of 
the two distinct spherically symmetric regions glued together. We will show 
that the gluing a LTB patch to the background FRW is in general not 
possible except for $\Sigma$ being a singular hypersurface carrying energy and 
momentum. To this end we write down the appropriate Israel junction equation 
on $\Sigma$\cite{khak}. It reads:
\begin{equation}
\epsilon_{in}\sqrt{1+\left( \frac{dR}{d\tau}\right)^{2}-
\frac{8\pi{\overline{\rho}}_b}{3}R^{2}} -
\epsilon_{out}\sqrt{1+\left( \frac{dR}{d\tau}\right)^{2}
-\frac{8\pi\rho_{b}}{3}R^{2}}\stackrel{\Sigma}{=}4\pi\sigma R ,
\end{equation}
where $\stackrel{\Sigma}{=}$ means that all functions on both sides
of the equality are evaluated on $\Sigma$, $\tau$ is the proper time 
of the comoving observer on $\Sigma$, $\sigma$ is the surface energy 
density of the boundary $\Sigma$, $\rho_b$ is the density at the boundary
being just a function of time and equal to the density of the background 
FRW universe, and ${\overline{\rho}}_b$ is the LTB mean density defined by the Eq. (4)
evaluated at the boundary $\Sigma$, i.e. the mean density of the local patch. 
The sign functions are fixed according to the convention 
$\epsilon_{in}(\epsilon_{out})= +1$ for $R$ increasing in the outward normal 
direction to $\Sigma$, while $\epsilon_{in}(\epsilon_{out})= -1$ for 
decreasing $R$. For the case we are considering with flat FRW metric, it 
can be shown that\cite{khak, sakai}
\begin{eqnarray}
\epsilon_{out}=sgn \left( 1 + v_{b}H_{b}R_b\right) ,
\end{eqnarray}
where $H_b$ is the Hubble parameter of the bulk, and $v_b$ is the radial peculiar 
velocity of $\Sigma$ relative to the bulk.  

\subsection{Constraints from the junction}

Now, without going into the detail discussion(see \cite{khak}), we may easily 
infer from the junction equation (10) that, in general, the matching is 
only possible for $\sigma \neq 0$, i.e. if a thin layer is formed on the
boundary of the mass condensation where our local patch joins the background
FRW. This is a mathematical possibility not observed yet, so we are going 
to discard it. The only exception is the case where $\epsilon_{in} = 
\epsilon_{out}$ and
\begin{equation}
\overline{\rho}_b : \stackrel{\Sigma}{=} \overline{\rho} = \rho_b,   
\end{equation}
 We, therefore, are left with the only case imposed by the dynamics of 
Einstein equations in which the mean density of our local patch is 
exactly equal to the density of the background FRW universe: a desired 
exact dynamical result reflecting the validity of the cosmological 
principle at large, contrary to the concerns of many authors assuming 
an underdense LTB region\cite{alnes, celerier, bolejko}. This fact can
be seen as a concrete example of the integral constraint in perturbing 
an energy-momentum tensor seen first by Traschen\cite{traschen}. Each 
nucleated patch within the FRW universe have the same average mass density 
as the bulk. Being distributed statistically, the patches does not have to 
destroy the homogeneity of the bulk. The total mass in a local patch, being
equal to the background density times the volume of the patch, is distributed
individually due to its self-gravity, leading to overdense structures and 
voids to compensate it. Assuming again the matter inside each patch to be 
smoothed out in the form of an inhomogeneous cosmic fluid, we expect it
to be overdense at the center decreasing smoothly to an underdense compensation 
region, a void, up to the point of matching to the background. We, therefore, have to expect voids around us, as it is indicated in different observations\cite{phillips}. Other cases is, however, possible depending on the fuctional form of $t_n$, being only constrained by the mean density. However, more general cases are conceivable, such as elliptic and hyporbolic cases in which $E(r) \not= 0$, even if the background is a flat FRW, which are outside the scope of this paper.\\ 
The equality of both sign functions is also an astrophysically trivial result. 
We know already from the technology of gluing manifolds\cite{khak, khak 01} 
that the sign functions are, for static metrics, related to the topology 
of the matching. In the case of non-static metrics, like FRW and LTB, the 
interpretation  is more complicated. Fortunately we are left with only one 
relatively trivial choice $\epsilon_{out} = \epsilon_{in} = +1$. In fact,
the case $-1$ is also possible, but it can easily be seen that it is isometric 
to the case where both sign functions are positive.  Within our model of an 
expanding FRW background from which a matter sphere is removed and replaced 
by a part of a LTB metric everything is topologically simple and is translated 
in the mathematical language as the positivity of both sign functions. This, 
again, is a desired astrophysical result coming out of the dynamics of 
Einstein equations. It remains to check the condition (7), which can now be 
written in the form
\begin{eqnarray}
\left( 1 + v_{b}H_{b}R_b\right) > 0 ,
\end{eqnarray}
where all quantities are to be taken at the boundary $\Sigma$. For a sub-horizon local patch, as it is assumed in our model, the inequality is valid for all values of the peculiar velocity $v_b$. \\ 
We are, therefore, left with a structured FRW universe for which the mean 
density of each local patch is equal to the FRW bulk density. The density 
distribution within a patch must be such that the overdensity of structures
are compensated by voids. Of course, for the actual mass distribution, taking into account the fine structure of the patch including the substructures, we have to rely on the overall observations and the matter power spectrum\cite{zehavi, goodwin, dekel, kocevski}.

\section{Luminosity distance-red shift relation and its astrophysical consequences} 

The luminosity distance in an LTB universe has been considered in many 
papers\cite{partovi, mustapha, celerier}, and also in a perturbed FRW universe\cite{barrausse, bonvin}. We will follow the paper\cite{celerier} as 
it is most suited to our purpose of comparing to observational data.
 
\subsection{Luminosity distance for small $z$ values}
The luminosity distance $d_L$ is assumed to be an explicit function of the red shift $z$, remembering its implicit dependence on the parameters of the model. According to the recent observations we expect the dimming of the cosmological objects at values of $z \approx 0.5 < 1$\cite{riess 04}. Besides this, just to check the possibility of explaining the dimming of cosmological objects as a consequence of the local inhomogeneities, we restrict our calculation in this paper to small $z$ values.  Therefore, we may legitimately use the Taylor expansion
around $z = 0$:
\begin{equation}
d_{L}(z) = d_1 z + d_2 z^2 + d_3 z^3 + O(z^4).
\end{equation}
In the FRW universe the coefficients of the expansion are given by\cite{celerier}
\begin{eqnarray}
&&  d_1 = \frac{1}{H_0}  ,\\
&& d_2 = \frac{1}{4H_0} \left(2 - \Omega_m + 2\Omega_{\Lambda}\right ), \\
&& d_3 =  \frac{1}{8H_0}\left(-2\Omega_m -4\Omega_{\Lambda} 
-4\Omega_m \Omega_{\Lambda} + {\Omega_m}^2 + 4{\Omega_{\Lambda}}^2\right ), 
\end{eqnarray}
where $H_0$ is the present time Hubble parameter and $\Omega_m$ and ${\Omega_{\Lambda}}$ 
are the familiar mass density- and cosmological constant-density parameters. \\
A straightforward calculation along the familiar line in FRW universe yields the 
coefficients of expansion in Eq.(10) for a LTB flat model\cite{celerier}:  
\begin{eqnarray}
&&  d_1 = \frac{1}{H_0},  \\
&&  d_2 = \frac{1}{4H_0} \left(1 - 6\frac{t_n^{'}}
                  {(\rho_c)^{\frac{1}{3}} t^{\frac{2}{3}}}\right ), \\
&&  d_3 = \frac{1}{8H_0} \left(-1 + 4\frac{t_n^{'}}
                  {(\rho_c)^{\frac{1}{3}} t^{\frac{2}{3}}} + 7\frac{t_n^{'2}}
                  {(\rho_c)^{\frac{2}{3}} t^{\frac{4}{3}}}-10\frac{t_n^{''}}{(\rho_c)^{\frac{2}{3}} t^{\frac{1}{3}}}\right ).
\end{eqnarray}
Here we have introduced the new coordinate $r$ such that 
$M(r) = \frac{1}{6} \rho_c r^3$ with constant $\rho_c$ in contrast to $\rho_c(r)$ defined in Eq.(7), indicating the comoving density. Obviously all functions are evaluated at the present time of the observer. The Hubble parameter is defined as\cite{partovi, humphreys}   
\begin{equation}
H_0 = \frac{1}{d_1}
    = \left(\frac{\dot R^{'}}{R^{'}}\right )_0.  
\end{equation}
The similarity of $d_1$ in both the FRW and LTB case should not obscure the fact that in the case of FRW the Hubble function $H$ is homogeneous and independent of the space coordinates. In our structured FRW model, we have to take into account the radial dependence of the Hubble function, in addition to its time dependence, reflected also in the peculiar velocity at the boundary of our patch. Hence, in fitting the observational data to the structured FRW model one has to use a local value for the Hubble parameter, and differentiate it from its mean global value. \\
 It can be seen from the coefficients of the expansion of the luminosity 
distance in powers of the redshift $z$ that the conventional FRW universe 
may mimic the coarse-grained FRW without the dark energy term. 
In fact, a comparison with the corresponding FRW coefficients shows 
that the luminosity distance coefficients of the structured FRW goes 
over to those of the FRW if one sets
\begin{eqnarray}
&& \Omega_m =  1 + 5\frac{t_n^{'}}
                  {(\rho_c)^{\frac{1}{3}}t^{\frac{2}{3}}} + \frac{29}{4} \frac{t_n^{'2}}
                  {(\rho_c)^{\frac{2}{3}}t^{\frac{4}{3}}}  
                  + \frac{5}{2} \frac{t_n^{''}}{(6\pi\rho_c)^{\frac{2}{3}}t^{\frac{1}{3}}}, \\
&&  \Omega_{\Lambda} =  - \frac{1}{2} \frac{t_n^{'}}
                  {(\rho_c)^{\frac{1}{3}}t^{\frac{2}{3}}}  
                  + \frac{29}{8} \frac{t_n^{'2}} {(\rho_c)^{\frac{2}{3}}t^{\frac{4}{3}}}
                  + \frac{5}{4} \frac{t_n^{''}} {(\rho_c)^{\frac{2}{3}}t^{\frac{1}{3}}}. 
\end{eqnarray}
These correspondence equations make a comparison with the observational data easier, as 
everyone is accustomed to the FRW jargon. Let us take as an example the data of\cite{Perlmutter 99} 
in the form
\begin{equation}
0.8\Omega_m - 0.6\Omega_{\Lambda} = - 0.2 \pm 0.1.
\end{equation}
The standard interpretation of this result in a FRW universe is that there is a dark energy
$\Omega_{\Lambda} > 0$. Substituting from equations (19 and 20) into (21) 
we obtain for the corresponding interpretation in a structured FRW universe
\begin{equation}
 4.3 \frac{t_n^{'}}
                  {(\rho_c)^{\frac{1}{3}}t^{\frac{2}{3}}}  
                  + 3.625 \frac{t_n^{'2}} {(\rho_c)^{\frac{2}{3}} t^{\frac{4}{3}}}
                  + 1.25 \frac{t_n^{''}} {(\rho_c)^{\frac{2}{3}}t^{\frac{1}{3}}}
                   = -1 \pm 0.1.
\end{equation}
We may also take the result of the first year of the 5-year {\it Supernova Legacy 
Survey}(SNLS)\cite{astier}. According to this survey we have
\begin{equation}
\Omega_m - \Omega_{\Lambda} = - 0.49 \pm 0.12.
\end{equation}
Substituting from the Eqs. (19 and 20) leads to 

\begin{equation}
 5.5 \frac{t_n^{'}}
                  {(\rho_c)^{\frac{1}{3}}t^{\frac{2}{3}}}  
                  + 3.625 \frac{t_n^{'2}} {(\rho_c)^{\frac{2}{3}} t^{\frac{4}{3}}}
                  + 1.25 \frac{t_n^{''}} {(\rho_c)^{\frac{2}{3}}t^{\frac{1}{3}}}
                   = -1.49 \pm 0.12.
\end{equation}
We are now ready to check if the our structured FRW model may leads to an enlarged luminosity distance for cosmic objects.

\subsection{The off-center observer and the nucleation time}

According to the cosmological principle we may be located anywhere off the center of our local patch. Therefore, our luminosity distance-redshift relation should be based on an off-center position of the observer and the corresponding past null geodesics. The off-center geometry of the light cone for LTB metric may be found in\cite{humphreys, bmn}. Let us take the simplest case of a radial off-center observation, where the observer is located at the point $P$ defined by the fixed coordinate $r = r_P$, and the source is located such that the center of the patch, the observer, and the source are aligned in the $\theta = \pi/2$ plane. In this case, the luminosity distance is given with a relation similar of the case of a central observer except for the functions $R$ and its derivatives which are to be taken at the point $r_P$(equations(35,36) in the reference\cite{humphreys} with the angle $\psi = 0$). Therefore, the expansion by $z$ is now at $r = r_P$, and we have to look at the behavior of $t_n$ and its derivatives at this point.\\
Let us now look at the behavior of the nucleation time ${t_n}^{'}$. From the equation(5) we have
\begin{equation}
t - t_n = R^\frac{3}{2} (\frac{9M}{2})^{-\frac{1}{2}}.
\end{equation}
Differentiating with respect to $r$ yields
\begin{equation}\label{nt}
t_n^{'} = - 9\pi R^{'} R^\frac{7}{2}(\frac{9M}{2})^{-\frac{3}{2}}(\overline{\rho} - \rho),
\end{equation}
Now, we have seen that $R^{'} > 0$, to avoid shell crossing. Therefore, the sign of $t_n^{'}$ at any point is determined by the difference between the mean density up to the coordinate value $r$ to the density at $r$. Depending on our position within the patch, we may have  $\overline{\rho} - \rho > 0$ or $< 0$. Therefore, $t^{'}_{n}$ may be either negative or positive. To avoid singularity at $r = 0$ we assume $\overline{\rho} = \rho$ at the center which leads to $t^{'}_{n}(r = 0) = 0$. At our position within the local patch defined now by 
$z = 0$, however, we may assume $t^{'}_{n}(r = r_P) < 0$, i.e. the mean density of structures up to our position, $r = r_P$, is larger than the density at our position, which means we are in an underdense region. This is the value we have to plug in the above equations to compare the luminosity distance to the observational data.\\

As it is well known\cite{Perlmutter 99, Riess 98, celerier} to interpret the SNIa data we need to take into account at least up to the $d_3$ term in the expression for the luminosity distance. The $d_3$ consists of terms proportional to the derivatives of $t_n$ up the the second order. Now, the influence of the $d_3$ term in the Eqs.(26, 28) can be seen in all the three terms on the left hand side of these equations. In a first approximation, we will now neglect the effect of the second derivative of $t_n$. From the fact that for a homogeneous metric $R(r,t) = a(t).r$, we can easily see that
\begin{equation}
  \alpha := \frac{t_n^{'}}{(6\pi\rho_c)^{\frac{1}{3}}t^{\frac{2}{3}}} \sim \frac{d t_n}{d(ar)}.
\end{equation}
Therefore, the term $\alpha$ is approximately equal to the running of $t_n$ with 
respect to the physical-, or even luminosity-, distance. Now, we can write the 
Eq. (28) as a second order equation in $\alpha$, which has the acceptable solution 
$\alpha = - 0.35$ in accordance with the assumption of our position to be in an underdense region. Ignoring the term proportional to the second derivative in the 
Eq. (24) we obtain $\Omega_m = 0.14$ and $\Omega_{\Lambda} = 0.63$. Note that in 
the evaluation of the observational data (23) the mean Hubble value, and not 
the local $H_0$, which is needed for a LTB comparison, is used. These are rough 
approximations which show explicitly the effect of inhomogeneity. Depending on the actual mass power spectrum, the second derivative may also be negative, which will increase the term corresponding to the dark energy. Note that
\begin{equation}
\alpha = -\frac{1}{6}(1 - \Omega_m + 2 \Omega_\Lambda),
\end{equation}
easily derived from Eqs.(23, 24), gives an exact relation between 
$\Omega_m $, $\Omega_\Lambda$, and the first derivative of $t_n$, independent of  the second derivative of $t_n$ present in both $\Omega_m$ and $\Omega_\Lambda$. If we assume the constraint 
$\Omega_m + \Omega_\Lambda = 1$, and take the value of $\alpha = -0.35$, we arrive at 
$\Omega_m = 0.3$ and $\Omega_\Lambda = 0.7$, which fits well to the results reported 
in\cite{clocchiatti} processing the data from their first sample of 75 low red shift 
and 43 high red shift SNIa. \\
We see, therefore, that our toy model for a realistic structured FRW universe, taking into account the impact of the inhomogeneities in our local patch, leads to a dimming of distant objects without the use of a dark energy.

\section{Exact backreaction of our local inhomogeneous LTB patch}

 The traditional way of doing cosmology is to take the average of the matter distribution in
the universe and write down the Einstein equations for it, adding some symmetry requirement. One then solves the equations $G_{\mu \nu} = \langle T_{\mu \nu} \rangle$, assuming homogeneity and isotropy of the mass distribution as the underlying symmetry. Note that in doing this we are taking the average of the energy momentum tensor at a constant time in the comoving coordinates, otherwise we would not come along with a homogeneously distributed matter content of the universe. As far as the precision of the observations allow, we may go ahead with this simplification. The more exact equation, however, is
$\langle G_{\mu \nu} \rangle = \langle T_{\mu \nu} \rangle$. Calling the difference $G_{\mu \nu} - \langle G_{\mu \nu} \rangle = Q_{\mu \nu}$, one may write the correct equation as $G_{\mu \nu} = \langle T_{\mu \nu} \rangle + Q_{\mu \nu}$. The backreaction term $Q$ has so far been neglected in cosmology because of its smallness. Now that measuring $Q$ is within the range of observational capabilities we have to take it into account. Of course, the averaging process is neither trivial nor unambiguous, but let us see what is the effect of a volume averaging in a comoving coordinates as it is done in the case of FRW model universe. 

\subsection{Volume averaging in the local patch}

We intend to average the inhomogeneities within our patch to get again a homogeneous patch within the FRW background and look for differences between this smoothed out SFRW and the original FRW. The difference caused by the backreaction is not vanishing and may have observational effects. To this end we will use the averaging formalism, developed mainly by Thomas Buchert\cite{b, b03, b057, b05}, which can easily be adapted to our LTB patch, having the same mass as the the FRW sphere cut out of it. In this formalism the volume-average of any function $f(t,r)$ is defined by
\begin{equation}
\langle f\rangle \equiv {1 \over V_D} \int_D dV f,
\end{equation}
where $dV$ is the proper volume element of the 3-dimensional domain $D$ of the patch we are considering and $V_D$ is its volume. It has been shown\cite{be,b} that in such a mass preserving patch the space-volume average of any function $f(r,t)$ does not commute with its time derivative:
\begin{equation}\label{nc}
\langle f\rangle^{\cdot} - \langle \dot f\rangle = \langle f\theta\rangle -  
\langle f\rangle \langle \theta \rangle,
\end{equation}
where the expansion scalar $\theta$, being equal to the minus of the trace of the second fundamental form of the hypersurface $t = const.$, is now a function of $r$ and $t$. The right hand side trivially vanishes for a FRW universe because of the homogeneity. This fact has far-reaching consequences for observational cosmology in our non-homogeneous neighborhood. The variation of the Hubble function with respect to the red-shift is not so simple any more as in the simple case of FRW universe. This affects a lot of observational data processing which so far has been done assuming homogeneity of the universe. Depending on the smoothing width $\Delta z$, the bins, and the matter power spectrum there may be large effects due to the non-commutativity of the averaging process\cite{e}.\\
The averaged scale factor is defined using the volume of our patch $D$ by $a_D \equiv V(t)_D^{1 \over 3}$. Now it can be shown that\cite{b, be}
\begin{equation}
\theta_D \equiv \langle \theta\rangle \equiv  {\dot V \over V} = 
           3{\dot a_D \over a_D} = 3 H_D.
\end{equation}
where we have used the notation $\dot a_D \equiv {d\over dt}a_D$, and denoted the average Hubble function as $H_D$. Averaging over the local patch means we are taking it as an effective FRW patch. Therefore all the derived quantities should be based on the average value $a_D$. This is why we take the above definition for the mean Hubble parameter and not $\langle \frac{\dot a}{a} \rangle$, which is different from $\frac{{\dot a}_D}{a_D}$. A similar difference holds for the second derivative of $a$: 
\begin{equation} 
 \langle\frac{\ddot a}{a} \rangle \not = \frac{\langle \ddot a \rangle} 
{\langle a \rangle} \not = {\ddot a_D \over a_D}.
\end{equation}
Therefore, the definition of the averaged deceleration parameter is not without ambiguity, specially because there is no nice relation like (9) for the deceleration parameter. To choose the most appropriate definition, we make recourse to the fact that in the averaging process we are taking our patch to be homogeneous and FRW-like. Therefore, in averaging the redshift as a function $a$, we always encounter $a_D$ and its time derivatives $\dot a_D$ and $\ddot a_D$. This justifies the above definition of the mean Hubble parameter and motivates us to make the following definition for the deceleration parameter:  
\begin{equation}
q_D  = - \frac{{\ddot a}_D a_D}{{\dot a_D}^2} =  - \frac{{\ddot a}_D}{a_D}\frac{1}{H_D^2},
\end{equation}
as was done in the literature so far\cite{b, hs,nt,kmr,s}. Now, we are ready to take the average of the Einstein equations in our local patch to see how the mean field equations will look like and what are the differences to the simple FRW field equations. Buchert's backreaction term is defined by\cite{b, b05} 
\begin{eqnarray}
Q = \langle \sigma^2 \rangle - {1\over3}\langle (\theta - \langle \theta \rangle)^2 \rangle \\
 = \langle \sigma^2 \rangle - {1\over3}[\langle \theta^2\rangle - \theta_D^2],
\end{eqnarray}
where $\sigma$ is the shear scalar and $\theta$ is the expansion. 
Although $\theta_D$ and $H_D$ are proportional, $\langle \theta^2 \rangle$ and $\langle H^2 \rangle$ are not. Hence, the relations (30, 37) can not be written in terms of $H$, as was done in\cite{nt}. The averages of the Einstein equations using the Hamiltonian constraint and the Raychaudhuri equation, taking into account the subtleties of the observation just mentioned, is then written in the following form\cite{b, b05}:
\begin{eqnarray}
\big({\dot a_D \over a_D} \big)^2 = {1\over3} (\rho_b+ \Lambda + Q)  \\
\frac{\ddot a_D}{a_D} = -{1\over6}(\rho_b - 2\Lambda + 4Q),
\end{eqnarray}
where we have set $\langle \rho \rangle = \rho_b$, the density of the background FRW universe, as a result of the junction conditions reflected in the eq.(11), and added the cosmological term for completeness. Note that in the so-called Friedmann equation (38) the averaged Hubble parameter enters instead of the global background one $H_b$. The effect of the backreaction within the local patch is realized as an effective extra perfect fluid having the density $\rho_Q = \frac{Q}{4\pi G}$, and pressure $p_Q$, and the equation of state
\begin{eqnarray}
\rho_Q = p_Q. 
\end{eqnarray}
A positive Q would lead to an increased Hubble parameter relative to the background $H_b$. In fact we have $H_b^2 = H^2_D - {1\over 3}Q$. Therefore, the averaged Hubble parameter measured in our subhorizon local patch is bigger than the background global one\\

\subsection{Explicit value of Q and its interpretation}

The value of $Q$ is determined by the balance between the mean values of the shear and the term related to the mean values of the Hubble parameter and the expansion scalar in a complex manner depending of the running of the density and the nucleation time. Given this complex behavior of the backreaction term, let us study it for the simplest case of the nucleation time satisfying the necessary conditions in our neighborhood discussed in the last section. We then approximate $t_n$ in the following way:
\begin{equation}
t_n = t_0 -\frac{\tau }{L^2} r^2,
\end{equation}
where $L = r_b$ is the comoving radius of the patch. For $\tau > 0$ the above expansion satisfies all the necessary conditions to be fullfilled by $t_n$ at the center of the patch and in our observational vicinity.  We obtain the following expression for the backreaction:
\begin{equation}
Q = \frac{(-5.8t +1.4\tau) + (2\tau)^{-\frac{1}{2}}t^{3\over 2}\lbrack 10.1 
\arctan (\sqrt{\tau t^{-1}}) - 1.7\arctan 1.5(\sqrt{\tau t^{-1}})\rbrack}{4\tau(t + \tau)^2} 
\end{equation} 

At the onset of nucleation, i.e. $t - t_0 \ll \tau$, the effects of backreaction is negligible. However, for the late time $t - t_0 \gg \tau$ we obtain $Q \approx 0.1 \frac{1}{\tau}\frac{1}{t}$. This is to be compared with $\frac{1}{t^2}$ behavior of the matter density. This suggest that the dimming of the SNIa distances we have seen in the last section must be due to the late time increase of $Q$ relative to the mass density and its effect on the background Hubble parameter.

\section{Conclusion}
 
 The precision cosmology is already so far developed that we can not ignore any more the effect of
the local inhomogeneities on the global cosmology. On the other hand, the well established 
successes of standard FRW cosmology can not be abandoned so easily. The structured FRW model (SFRW) we are proposing is just a step further towards a more realistic model universe, and is in accordance with the cosmological principle. In fact it could be considered as an exact Swiss Cheese model, in which the cheese is in the holes: a FRW model in which there are local subhorizon inhomogeneous patches embedded homogeneously as an exact solution of the Einstein equations. Each local patch is approximated, therefore, by an inhomogeneous cosmic fluid represented by a LTB metric up to a radius $r_b$ embedded in a background homogeneous FRW universe. Each local inhomogeneous sphere is then glued to a FRW homogeneous universe from which a sphere of the same radius is removed. The dynamics of the Einstein equations leave only one possibility for such a matching which is astrophysically appealing: The patches consist of overdense and underdense compensating regions such that the mean density in each patch is equal to the background FRW density. Taking into account this junction condition, the luminosity distance-redshift relation shows a dimming of astrophysical objects relative to what may be inferred from a simple FRW universe.
We have analyzed the luminosity distance from a cosmic object to an on-center or off-center observe in such a SFRW universe and shown that a dimming of cosmic objects, which could mimic a dark energy, is the result of the inhomogeneity.  \\
We are used to average out the inhomogeneities within the universe and take a global FRW metric to represent it. Within the proposed SFRW universe we may also average out the inhomogeneities. The process of averaging is adopted to the familiar understanding that the local inhomogeneities in the universe at any time $t$ should be smoothed out. To that end we have used the volume averaging of the Einstein equations developed so far to see the effect of the backreaction within a homogeneous scenario. The consequence is a backreaction term that maybe interpreted in terms of a new effective energy momentum tensor, although it is just a geometric term modifying the Friedmann equations. Using this interpretation, one could say that in addition to the mean cosmic matter fluid, we have a backreaction fluid with a density and pressure which has a very peculiar behavior. The back reaction density  behaves as $1\over t$ at late times and leads to a reduced background Hubble parameter. \\
 Our simple model of a structured FRW universe shows how important 
it is in the era of precision cosmology to go beyond the model of a simple homogeneous 
fine-grained cosmic fluid and replace it by a scenario with coarse-grained local 
patches. The fact that the value of the mean local Hubble parameter, is not equal to 
the global one, and the collective peculiar velocity of the objects reflected 
in that of cosmic fluid are just some of the changes in the terminology of the 
universe models we have to incorporate in the interpretation of the 
astrophysical data. 
  
\section{Acknowledgments}
 It is a pleasure to thank members of the cosmology group at Sharif University of 
Technology, specially S. Khakshournia and S. Rahvar, the members of the cosmology 
group at McGill University, specially Robert Brandenberger, and E. W. Kolb for discussions
about the main idea of this work. My special thanks to McGill Physics department and 
Robert Brandenberger for the hospitality.


\begin{thebibliography}{99}

\bibitem{Peebles} P. J. E. Peebles, {\it The large-Scale Structure of the Universe}, Princeton
        University Press, Princeton, New Jersey, 1980.
\bibitem{Lahav} O. Lahav, {\it Proceeding of the Nato Advanced Study Institute on 
         Structure Formation in the Universe}, Edited by R. G. Crittenden and N. G. Turok
	 (Kluver Academic Publishers, Dordrecht 2001)Nato Science Series, Series C: 
	 565, 131; astro-ph/0001061.
\bibitem{Riess 98} A. G. Riess et. al., Astron. J., 116, 1009 (1998); astro-ph/ 9805201
\bibitem{Perlmutter 99} S. Perlmutter et. al., Astrophys.J., 517, 565 (1999); 
                        astro-ph/981.
\bibitem{Riess 04} A.G. Riess et. al., Astrophys. J., 607, 665 (2004).
\bibitem{Fillipenko 04} A. V. Fillipenko, astro-ph/0410609.
\bibitem{Kolb 05} E. W. Kolb, S. Matarrese, and A. Riotto, astro-ph/0506534.
\bibitem{Geshniz 02} G. Geshnizjani and R. Brandenberger, Phys. Rev. D66, 
         123507, 2002; gr-qc/0204074
\bibitem{bran 02} R. Brandenberger, hep-th/0210165.
\bibitem{Martineau 05} P. Martineau and R. Brandenberger, astro-ph/0510523.
\bibitem{KMNR 05} E. W. Kolb S. Matarrese, A. Notari, and A. Riotto, hep-th/0503117.
\bibitem{rasanen} S. Rasanen, astro-ph/0504005.
\bibitem{notari}, A. Notari, astro-ph/0503715.
\bibitem{KMR 05} E. W. Kolb, S. Matarrese, and A. Riotto, astro-ph/0511073.
\bibitem{Geshniz 05} G. Geshnizjani, D. J. H. Chung, and N. Afshordi, astro-ph/0503553.
\bibitem{hirata 05} Ch. M. Hirata and U. Seljak, astro-ph/0503582.
\bibitem{Flanagan 05} E. E. Flanagan, astro-ph/0503202.
\bibitem{partovi} M. H. Partovi and B. Mashhoon, Astrophys. J., 276, 4, 1884.
\bibitem{celerier} M. Celerier, A\&A, 353, 63, 2000; astro-ph/0512103.
\bibitem{LTB} G. Lemaiter, Ann. soc. Sci. Bruxelles Ser.1, A53, 51, 1933; 
              R. C. Tolman,Proc. Nat1. Acad. Sci. U.S.A. 20,410, 1934; 
              H. Bondi, Mon. Not. R. Astron. Soc., 107, 343, 1947).
\bibitem{Tomita 00} K. Tomita, Astrophys. J., 529, 38, 2000; astro-ph/0005031.
\bibitem{Tomita 01} K. Tomita, Prog. Theor. Phys. 106, No. 5, 2001.
\bibitem{giovanni} M. Giovannini, hep-th/0505222.
\bibitem{wiltshire 05} D. L. Wiltshire, gr-qc/0503099.
\bibitem{moffat} J. M. Moffat, astro-ph/0502110 and 0505326.
\bibitem{alnes} H. Alnes, M. Amarzguioui, and O. Gron, astro-ph/0512006.
\bibitem{bolejko} K. Bolejko, astro-ph/0512103.
\bibitem{khak} S. Khakshournia and R. Mansouri, Phys. Rev. D65, 027302, 2003;\\ gr-qc/0307023. 
\bibitem{riess 04} A. G. Riess et. al., Astrophys. J., 607, 665-687, 2004; astro-ph/05402512.

\bibitem{hellaby} C. Hellaby and A. Krasinsky, gr-qc/0510093 (to be published in Phys. Rev. D)
\bibitem{silk}  D. Olson and J. Silk, Astrophys. J., 233, 395, 1979.
\bibitem{mansouri} R. Mansouri, Ann. Inst. Henri Poincare, XXVII, 175, 1977.
\bibitem{sakai} N. Sakai and K. I. Maeda, Phys. Rev. D50, 5425, 1994.
\bibitem{traschen} J.Traschen, Phys. Rev. D31, 283, 1985.
\bibitem{phillips} L. A. Phillips and E. L. Turner, astro-ph/9802352.
\bibitem{khak 01} S. Khakshournia and R. Mansouri, Gravitat. \& Cosmology, 7, 261, 2001.  
\bibitem{hui} L. Hui and P. B. Greene, astro-ph/0512159.
\bibitem{zehavi} I. Zehavi et. al. Astrophys. J. 503, 483, 1998; astro-ph/9802252.
\bibitem{goodwin} S.P. Goodwin, P. A. Thomas, A. J. Barber, J. Gibbon, and L. I. Onuora, astro-ph/9906187.
\bibitem{dekel} A. Dekel, ARA\&A, 32, 371, 1994.
\bibitem{kocevski} D. D. Kocevski, H. Ebeling, Ch. R. Mullis, and R. Brent Tully, astro -ph/0512321.
\bibitem{mustapha} N. Mustapha, C. Hellaby, and G. F. R. Ellis, MNRAS, 292, 817, 1997; 
                   gr-qc/9808079	
\bibitem{barrausse} E. Barausse, S. Matarrese, A. Riotto, Phys. Rev. D71,
                    063537, 2005; astro-ph/0501152.
\bibitem{bonvin} C. Bonvine, Ruth Durrer, and M. A. Gasparini, astro -ph/0511183.
\bibitem{humphreys} N.P. Humphreys, R. Maartens, and D. R. Matravers, Astrophys. J., 477, 47, 1997.
\bibitem{astier} P. Astier et. al., astro-ph/0510447.		    
\bibitem{bmn} Thirthabir Biswas, Reza Mansouri, and Alessio Notari, Nonliner Structure Formation and apparent Acceleration without Dark Energy, work in progress.
\bibitem{clocchiatti} A. Clocchiatti et. al. astro-ph/0510155.
\bibitem{b} T. Buchert, Gen. Rel. Grav. 32, 105, 2000 and 33, 1381, 2001;                  gr-qc/9906015 and gr-qc/0102049;                                 
\bibitem{b03} T. Buchert, Phys. Rev. Lett. 90, 031101, 2003; gr-qc/0210045; 
\bibitem{b057} T. Buchert, Class. Quant. Grav., 22, L113, 2005; gr-qc/0507028; 
\bibitem{b05} T. Buchert, Class. Quant. Grav., 23, 817, 2006; gr-qc/0509124.
\bibitem{e}D. J. Eisenstein et.al., Astrophys.J. 633, 560, 2005; astro-ph/0501171.
\bibitem{be} Thomas Buchert and Juergen Ehlers, A\&A, 320, 1, 1997.
\bibitem{nt} Y. Nambu and M. Tanimoto, gr-qc/0507057.

\end{thebibliography}
\end{document}